\documentclass[%
reprint,
showpacs,preprintnumbers,
nofootinbib,
nobibnotes,
amsmath,amssymb,
prb,
floatfix,
longbibliography
]{revtex4-2}

\usepackage{graphicx}
\usepackage{dcolumn}
\usepackage{bm}
\usepackage{colortbl}
\usepackage[unicode=true,colorlinks=true,citecolor=blue,urlcolor=blue]{hyperref}
\usepackage{braket}
\usepackage[normalem]{ulem}
\usepackage{enumitem}

\graphicspath{{img/}}

\newcommand{\divv}{\mathop{\rm div}}

\begin{document}

\title{Viscous hydrodynamics of excitons in van der Waals heterostructures}

\author{V.~N.~Mantsevich}
\affiliation{Lomonosov Moscow State University, 119991 Moscow, Russia}
\author{M.~M.~Glazov}
\affiliation{Ioffe Institute, 194021 St. Petersburg, Russia}

\begin{abstract}
 Excitons in semiconductors can form a variety of collective states leading to different regimes of exciton propagation. Here we theoretically demonstrate the possibility to reach the viscous hydrodynamic -- liquid-like -- regime of exciton propagation in two-dimensional materials, focusing on the mono- and bi-layers of transition metal dichalcogenides. This regime can be realized where the exciton-exciton collisions dominate over exciton-phonon and disorder scattering. We have derived the hydrodynamic-like set of equations describing viscous flow of interacting excitons based on the Boltzmann kinetic equation for the exciton distribution function. A comparison of various exciton propagation regimes including diffusive, viscous hydrodynamic, and superfluid regime is presented. Conditions which allow one to observe the hydrodynamic regime of exciton transport, and the role of material are discussed.  
 \end{abstract}

\maketitle{}

\section{Introduction}\label{sec:intro}

Two-dimensional (2D) transition-metal dichalcogenides form a unique platform to study physics of interacting systems~\cite{Kolobov2016book,deotare20232d}. Their optical response is dominated by the Coulomb bound quasiparticles, excitons, as well as more complicated quasiparticles such as trions or biexcitons~\cite{Ivchenko2005}. Large, on the order of hundreds of millielectronvolts exciton binding energies allow one to observe and study excitons even at the room temperatures~\cite{RevModPhys.90.021001}.  Excitons not only strongly contribute to the optical properties of 2D semiconductors and van der Waal heterostructures on their basis~\cite{Geim:2013aa}, but are also involved in the energy transport as they are free to move across the material plain \cite{Hillmer1988,Steininger1996,Rapaport2004,Kumar2014,Kato2016}. Moreover, strongly bound and long-lived propagating excitons can act as information carriers within a semiconductor and are highly promising for potential applications in chip-scale optical processing systems \cite{Baldo2009,Causin2022}. 

A combination of fundamental physics and potential applications makes 2D exciton transport a highly topical field of research nowadays~\cite{Malic:2023aa,Chernikov:2023ab}.
Experimental studies of exciton transport in 2D materials revealed diffusive propagation across hundreds of nanometers at ambient conditions \cite{Kumar2014,Yuan2017}. The key parameters of exciton propagation such as their diffusion coefficient and diffusion length depend on the disorder \cite{Vlaming2013,Kurilovich2020,Kurilovich2022,Lee2015}, exciton-phonon interaction~\cite{Glazov2019,Zipfel2020,Glazov2020}, presence of free charge carriers~\cite{Wagner2023} and are strongly influenced by inter-particle interactions~\cite{Uddin2020,Cheng2021,Wagner2023}. Theoretical analysis of exciton transport is usually based on well-known approaches to transport phenomena in semiconductor systems \cite{Shklovskii1984,Gantmakher1987} taking into account the specifics of excitons. The most widely used models for exciton transport analysis are semiclassical models based on Boltzmann kinetic equation \cite{Zipfel2020,Choi2023}. Semiclassical approach can be extended to take into account quantum effects by using the approach based on the diagram technique \cite{Glazov2020}. Alternative approaches are based on microscopic picture, such as tight-binding models \cite{Kenkre1983,Heijs2005} or kinetic Monte Carlo calculations \cite{Akselrod2014_1,Miyazaki2012}. Among the most intriguing results concerning exciton transport in diffusion regime in 2D materials is the appearance of negative effective diffusivity discussed in \cite{Wietek2024,Kurilovich2023}. Beyond semiclassical picture, the quantum mechanical approaches based on the wave packet propagation are also developed revealing intricate features of excitonic bands~\cite{Qiu:2021to}.

While diffusive propagation of excitons remains the most studied and widespread transport regime in 2D semiconductors~\cite{Chernikov:2023ab} other scenarios are, generally, possible and deserve detailed study. Particularly interesting propagation regimes are realized where excitons behave collectively. The most prominent example is the superfluid propagation of excitons discussed in the literature~\cite{Fogler2014}. The transition to superfluidity is related to the formation of degenerate Bose-gas of excitons accompanied by the Berezinksii-Kosterlitz-Thouless transition, see Ref.~\cite{GlazovSuris_2021} and references therein. Hence, at  low temperatures and relatively high densities the excitons should behave as inviscid fluid. Recently signatures of  an inviscid propagations have been reported for a MoS$_2$ monolayer~\cite{Aguila2023}. The experimental results are consistent with the non-viscous fluid scenario, however, the ultrahigh speed of exciton propagation, being several percents of speed of light call for further discussion~\cite{Glazov:2023aa,glazov2024ultrafastexcitontransportvan}. All this motivates further studies of non-diffusive exciton propagation in atomically thin semiconductors.

Here we show that, in addition to a superfluid, the excitons can form a viscous liquid. It happens in the case where excitons are still non-degenerate, but exciton-exciton scattering rate exceeds the exciton-phonon and exciton-disorder scattering rates. In such a situation, the exciton-exciton collisions determine the viscosity of the fluid. Similar viscous hydrodynamic transport regime is now actively studied for electrons in semiconductors and semiconductor nanosystems both theoretically and experimentally~\cite{Gurzhi_1968,PhysRevLett.106.256804,PhysRevLett.117.166601,PhysRevB.51.13389,Bandurin1055,Crossno:2016aa,Moll1061,Levitov:2016aa,Gusev:2018tg,PhysRevLett.128.136801,Narozhny:2022ud}.  

Based on the kinetic equation for the exciton distribution function we derive the analogue of Navier-Stokes equations for exciton propagation and microscopically determine their viscosity. We analyze the range of parameters where such hydrodynamic behavior can be realized, highlighting its particular importance for bilayer 2D semiconductors.
The paper is organized as follows. After introduction (Sec.~\ref{sec:intro}) we derive basic equations for viscous hydrodynamics in excitons in Sec.~\ref{sec:hydro}. Next, in Sec.~\ref{sec:phase} we present exciton propagation regimes and derive the criteria to observe the viscous hydrodynamics in transition metal dichalcogenide based mono- and bilayers. Section~\ref{sec:transport} contains the results of calculations of exciton propagation in various regimes. Finally, the conclusion is given in Sec.~\ref{sec:conclusion}.

\section{Hydrodynamic equations and exciton viscosity}\label{sec:hydro}

This section presents derivation of the key model equations of our work: exciton hydrodynamic equations. To that end we use the kinetic (Boltzmann) equation for exciton distribution function in the presence of exciton-exciton interactions. Under assumption that the exciton-exciton scattering processes dominate over other scattering mechanisms we derive a set of Navier-Stokes like equations for excitonic fluid and evaluate the viscosity caused by the exciton-exciton collisions. The derived equations are used in the following sections to model exciton propagation in various geometries and analyze the regimes of excitonic transport. 
As a model system we consider excitons in a 2D layer or bilayer of transition metal dichalcogenides, but our basic results are valid for any two-dimensional excitonic system.

\subsection{Hydrodynamic equations}

We consider 2D excitons and describe the exciton ensemble via the time $t$, in-plane coordinate $\mathbf r = (x,y)$, and wavevector $\mathbf k$-dependent distribution function $f_{\mathbf k}(\mathbf r, t)$. Here for brevity we consider only one type of excitons, the generalization of the developed model to account for several types of excitons, e.g., bright and dark ones, or intra- and interlayer- ones, or spin-up and spin-down excitons is straightforward. We consider isotropic parabolic dispersion of excitons in the form 
\begin{equation}
\label{dispersion}
E_k = \frac{\hbar^2 k^2}{2m},
\end{equation}
where $m$ is the exciton translational motion mass and $\hbar$ is the reduced Planck constant. Let 
\begin{equation}
\label{density}
n(\mathbf r,t) = \sum_{\mathbf k} f_{\mathbf k}(\mathbf r,t)
\end{equation}
be the exciton density. The normalization area is set to be unity hereafter and the summation over degenerate spin or valley states is omitted here, it can be included in $\sum_{\mathbf k}$.

Under the principal assumptions
\begin{subequations}
\label{assumptions}
\begin{align}
\frac{mk_B T}{\hbar^2 n} \gg 1, \label{non:deg}\\
na_B^2 \ll 1, \label{below:mott}\\
\frac{k_B T}{E_B} \ll 1, \label{below:EB}\\
\frac{k_B T\tau}{\hbar} \gg 1, \label{no:quantum}
\end{align}
\end{subequations}
where $a_B$ is the exciton Bohr radius, $E_B$ is the exciton binding energy, $\tau$ is the characteristic scattering time, $T$ is the temperature and $k_B$ is the Boltzmann constant, the exciton distribution function obeys the kinetic equation 
\begin{eqnarray}
\frac{\partial f_{\mathbf k}}{\partial t}+\textbf{v}_{\mathbf k}\frac{\partial f_{\mathbf k}}{\partial \textbf{r}}+\frac{1}{\hbar}\frac{\partial U}{\partial \textbf{r}}\frac{\partial f_{\mathbf k}}{\partial \textbf{k}}=Q_{xx}\{f_{\mathbf k}\}+Q_p\{f_{\mathbf k}\},
\label{eq:Boltzmann_0}
\end{eqnarray}
In kinetic Eq.~\eqref{eq:Boltzmann_0} $\mathbf v_{\mathbf k} = \hbar^{-1} \partial E_k/\partial \mathbf k$ is the exciton group velocity, $U\equiv U(\mathbf r,t)$ is the exciton potential energy owing to exciton-exciton interactions and external fields, $Q_{xx}\{f_{k}\}$ and $Q_p\{f_{\mathbf k}\}$ are the collision integrals describing, respectively, the exciton-exciton scattering and the exciton scattering by static impurities and phonons, their explicit form is specified below. The arguments $(\mathbf r, t)$ of the distribution function and potential energy are omitted for brevity.
Note that Eq.~\eqref{non:deg} allows us to consider excitons as non-degenerate, otherwise exciton-exciton interactions induce superfluidity in the system, see~\cite{Fogler2014,GlazovSuris_2021} and references therein. The condition~\eqref{below:mott} guarantees that the exciton density is sufficiently small (below the Mott transition density) so the exciton dissociation is negligible, similarly, under condition~\eqref{below:EB} thermal dissociation of excitons can be disregarded. The condition~\eqref{no:quantum} allows us to disregard quantum correlations between different states. 

Generally, Eq.~\eqref{eq:Boltzmann_0} should be supplemented by the generation and recombination terms. However, the exciton lifetime is by far longer than the exciton collision times~\cite{Chernikov:2023ab}, hence, hereafter we consider excitons as everlasting particles. The finite lifetime can be readily added in the developed formalism, see below. 

To derive hydrodynamic equations we obtain, by summing Eq.~\eqref{eq:Boltzmann_0} over $\mathbf k$ the continuity equation in the form
\begin{equation}
\label{cont}
\frac{\partial n}{\partial t}+\divv(n \mathbf{V})=0,
\end{equation}
where 
\begin{equation}
\label{hydro:V}
\mathbf V \equiv \mathbf V(\mathbf r, t) = \frac{1}{n(\mathbf r,t)}\sum_{\mathbf k} \mathbf v_{\mathbf k} f_{\mathbf k}(\mathbf r,t),
\end{equation}
is the exciton hydrodynamic velocity that corresponds to motion of the exciton fluid as a whole. In the presence of generation and finite lifetime of excitons the right hand side of Eq.~\eqref{cont} should be written as $G(\mathbf r,t)-R[n(\mathbf r,t)]$, i.e., as a difference of the generation and recombination rates.

To derive the equation for the hydrodynamic velocity $\mathbf V$ we multiply the kinetic equation by $\mathbf v_{\mathbf k}$ and sum over $\mathbf k$. First, let us neglect the momentum relaxation processes setting $Q_p\{f_{\mathbf k}\} =0$. Hence, the distribution function is controlled by exciton-exciton collisions and, naturally, takes the quasiequilibrium form~\cite{Gantmakher1987,ll10_eng}
\begin{multline}
\label{equil}
f_{\mathbf k} = f_{\mathbf k}^0 + \delta f_{\mathbf k}, \quad f_{\mathbf k}^{0}=\exp{\left(\frac{\mu-E_{|\mathbf k - m \mathbf V/\hbar|}}{k_B T}\right)}\\
\approx \exp{\left(\frac{\mu-E_{k}+ \hbar\mathbf k\cdot \mathbf V}{k_BT}\right)},
\end{multline}
with $\mu$ being chemical potential.
The Boltzmann distribution in Eq.~\eqref{equil} is shifted in the $\mathbf k$-space because excitons flow with the velocity $\mathbf V$, Eq.~\eqref{hydro:V}. The latter approximate equality holds for $V \ll \sqrt{k_B T/m}$. The distribution in the form~\eqref{equil} nullifies the exciton-exciton collision integral $Q_{xx}\{f^0_{\mathbf k}\} =0$, and a small correction $\delta f_{\mathbf k}$ appears owing to gradients of the macroscopic parameters in $f_{\mathbf k}^0$: density, temperature, and velocity. As a result we obtain from Eq.~\eqref{eq:Boltzmann_0}
\begin{equation}
\label{flux}
\frac{\partial (n\mathbf V)}{\partial t} = -\divv \hat{\Pi},
\end{equation}
where the second-rank momentum flux density tensor $\hat{\Pi}$ has the following Cartesian components ($\alpha,\beta=x$ or $y$)
\begin{eqnarray}
\Pi_{\alpha\beta}=P\delta_{\alpha\beta} + n V_\alpha V_\beta-\sigma_{\alpha\beta}.
\label{Pi:tensor:def}
\end{eqnarray}
In what follows we omit quadratic in $\mathbf V$ terms assuming that velocity of excitons is small.
Here $P$ is the pressure associated with the kinetic and potential energies of excitons and $\sigma_{\alpha\beta}$ is the viscous stress tensor. They read, respectively,
\begin{subequations}
\label{eq:hydro_system_2}
\begin{equation}
\label{pressure}
P=n\frac{k_{B}T}{m}+ \frac{n}{m} U_{\rm ext} + \frac{n^2}{2m}U_0,
\end{equation}
and
\begin{eqnarray}
\label{sigma:0}
\sigma_{\alpha\beta}=-\frac{\hbar^2}{m}\sum_{\mathbf k} \left(k_\alpha k_\beta-\frac{1}{2}\delta_{\alpha\beta}k^{2}\right)\delta f_{\mathbf k}.
\end{eqnarray}
or
\begin{eqnarray}
\label{sigma}
\sigma_{\alpha\beta}=\eta\left(\frac{\partial V_{\alpha}}{\partial x_{\beta}}+\frac{\partial V_{\beta}}{\partial x_{\alpha}}-\delta_{\alpha\beta}\divv\mathbf V\right).
\end{eqnarray}
\end{subequations}
Here $\eta$ is the exciton viscosity (see Sec.~\ref{subsec:visc}), $U_{\rm ext}$ is the external potential acting on excitons and $U_0$ is the exciton-exciton interaction constant. We assumed pair contact interactions between the particles described by the interaction potential energy 
\begin{equation}
\label{V:inter}
V(\mathbf r - \mathbf r') = U_0\delta(\mathbf r - \mathbf r'),
\end{equation} 
such form of the interaction in justified below in Secs.~\ref{subsec:mono} and \ref{subsec:bi}, we only mention here that the hydrodynamic description is valid on the length scales that exceed by far any microscopic length in the system including the interaction potential radius~\cite{Irving:1950aa}. Note that the contribution in Eq.~\eqref{sigma} is traceless. The second viscosity contribution to the stress tensor $\propto \delta_{\alpha\beta} \divv{\mathbf V}$ vanishes in our system similarly to the case of a monoatomic gas~\cite{ll10_eng}.

Before proceeding to calculation of the exciton viscosity we include the momentum relaxation processes into account. To that end we represent 
\begin{equation}
\label{momentum:rel}
\sum_{\bm k} \mathbf v_{\mathbf k} Q_p\{f^0_{\mathbf k}\} = - \frac{n\mathbf V}{\tau_p},
\end{equation}
where $\tau_p$ is the momentum relaxation time. Indeed, the collision integral $Q_p\{f^0_{\mathbf k}\}$ does not vanish because $f^0_{\mathbf k}$ has an anisotropic part $\propto \mathbf k \cdot \mathbf V$. One can readily show that for elastic or quasi-elastic collisions the momentum relaxation rate, $1/\tau_p$, in Eq.~\eqref{momentum:rel} can be expressed as~\cite{Gantmakher1987}
\begin{equation}
\label{tau:p:aver}
\frac{1}{\tau_p} = \frac{\sum_{\mathbf k} \tau_p^{-1}(E_k) E_k \exp{(-E_k/k_B T)}}{ \sum_{\mathbf k} E_k \exp{(-E_k/k_B T)}}
\end{equation}
via the energy-dependent momentum relaxation rate $\tau_p^{-1}(E_k) \equiv (2\pi/\hbar) \sum_{\mathbf k'} |M_{\mathbf k'-\mathbf k}|^2 (1-\cos{\vartheta_{\mathbf k, \mathbf k'}})\delta(E_k - E_k')$. Here $\vartheta_{\mathbf k, \mathbf k'} = \angle{\mathbf k, \mathbf k'}$ and $M_{\mathbf k'-\mathbf k}$ is the matrix element of exciton-defect or exciton-phonon interaction.\footnote{If exciton-exciton scattering were negligible the exciton diffusion coefficient is determined by $$\tau_1 = \left(\sum_{\mathbf k} \tau_p(E_k) {E_k} e^{-E_k/k_B T}\right)/\left( \sum_{\mathbf k} {E_k} e^{-E_k/k_B T}\right),$$ i.e., where the averaging is performed of $\tau_p(E_k)$ rather than of $1/\tau_p(E_k)$ in Eq.~\eqref{tau:p:aver}.}

Finally, hydrodynamic equations for viscous flow of exciton read
\begin{subequations}
\label{excitons:hydro}
\begin{equation}
\label{cont:fin}
\frac{\partial n}{\partial t}+\divv(n \mathbf{V})=G(\mathbf r,t)-R[n(\mathbf r,t)],
\end{equation}
where the continuity equation~\eqref{cont:fin} accounts for (slow) generation and recombination processes and
\begin{equation}
\label{flux:fin}
\frac{\partial (n\mathbf V)}{\partial t} + \frac{n\mathbf V}{\tau_p} = -\divv \hat{\Pi},
\end{equation}
\end{subequations}
accounts for the momentum relaxation. Equation~\eqref{flux:fin} can be recast in a form similar to the Navier-Stokes equation~\cite{ll6_eng}
\begin{multline}
\label{NS}
\frac{\partial(nV_{\alpha})}{\partial t}+\frac{nV_{\alpha}}{\tau_p} 
\\
=-\frac{\partial P}{\partial x_\alpha} +\eta\Delta V_{\alpha}
+\frac{\partial \eta}{\partial x_{\beta}}\left(\frac{\partial V_{\alpha}}{\partial x_{\beta}}+\frac{\partial V_{\beta}}{\partial x_{\alpha}}-\delta_{\alpha\beta}\frac{\partial V_{\gamma}}{\partial x_{\gamma}}\right).
\end{multline}
that takes into account specific features of exciton fluid, namely, generation-recombination processes, momentum relaxation, and its compressibility. The term with the pressure gradient can be conveniently recast
\[
\frac{\partial P}{\partial \mathbf r} = -\frac{k_B T}{m} \frac{\partial n(\mathbf r,t)}{\partial \mathbf r} - \frac{n}{m}\frac{\partial U}{\partial \mathbf r},
\]
i.e., as a generalized force density acting on the excitons.

\subsection{Viscosity coefficient}\label{subsec:visc}

Let us now derive the viscosity coefficient $\eta$. To that end we need to determine a small correction to the distribution function, $\delta f_{\mathbf k}$, in Eq.~\eqref{equil} that appears in the presence of the velocity gradient. The linearized kinetic Eq.~\eqref{eq:Boltzmann_0} neglecting the momentum relaxation processes\footnote{If momentum relaxation processes dominate over exciton-exciton collisions the hydrodynamic approach becomes, strictly speaking, invalid. However, for small gradients general Eqs.~\eqref{excitons:hydro} hold true, see Sec.~\ref{sec:transport} for details.} reads
\begin{eqnarray}
\textbf{v}_{\mathbf k}\frac{\partial f_{\mathbf k}^{0}}{\partial \textbf{r}}=Q_{xx}\{f_{\mathbf k}\},
\end{eqnarray}
where we neglected the time derivative of the distribution function as compared with the contribution of collision integral.\footnote{As already mentioned, the hydrodynamic approach is valid on the time scales that strongly exceed microscopic collision times. Hence, for determining $\delta f_{\mathbf k}$ the problem can be considered static.} Making use of the explicit form of the quasiequilibrium distribution~\eqref{equil}, one can get an equation for non-equilibrium correction of the distribution function $\delta f_{\mathbf k}$
\begin{eqnarray}
\frac{f_{k}^{0}}{k_BT}mv_{k,\alpha}v_{k,\beta}V_{\alpha\beta}=Q\{\delta f_{k}\},
\label{eq:delta_f}
\end{eqnarray}
where the shift of the quasiequilibrium distribution can be disregarded and we introduced notation $V_{\alpha\beta}={\partial V_{\beta}}/{\partial x_{\alpha}}$ for the tensor of hydrodynamic velocity gradients. It follows from the microscopic analysis presented in Appendix~\ref{app:xx} that
\begin{eqnarray}
\label{dfk:temp}
\delta f_{\mathbf k}=-\frac{\hbar^2 f_{k}^{0}}{m k_BT}\tau_{xx}k_\alpha k_\beta V_{\alpha\beta},
\label{eq:delta_f_final}
\end{eqnarray}
where $\tau_{xx}$ is the relaxation time of the second angular harmonics of the exciton distribution function
\begin{eqnarray}
\frac{1}{\tau_{xx}}= n \frac{m U_{0}^{2}}{\hbar^{3}}.
\label{eq:ex_ex_scattering}
\end{eqnarray}
Here, as in Eq.~\eqref{pressure}, $U_0$ is the exciton-exciton interaction constant. Substituting $\delta f_{\mathbf k}$ from Eq.~\eqref{eq:delta_f_final} into Eq.~\eqref{sigma:0} we arrive at Eq.~\eqref{sigma} for the viscous stress tensor with the viscosity coefficient
\begin{equation}
\label{eta:fin}
\eta=\frac{n}{m}\tau_{xx}k_BT.
\end{equation}
Equations~\eqref{eq:ex_ex_scattering} and \eqref{eta:fin} provide microscopic expression for the key parameter of exciton hydrodynamics, their viscosity.

\begin{figure*}[t]
\includegraphics[width=0.99\linewidth]{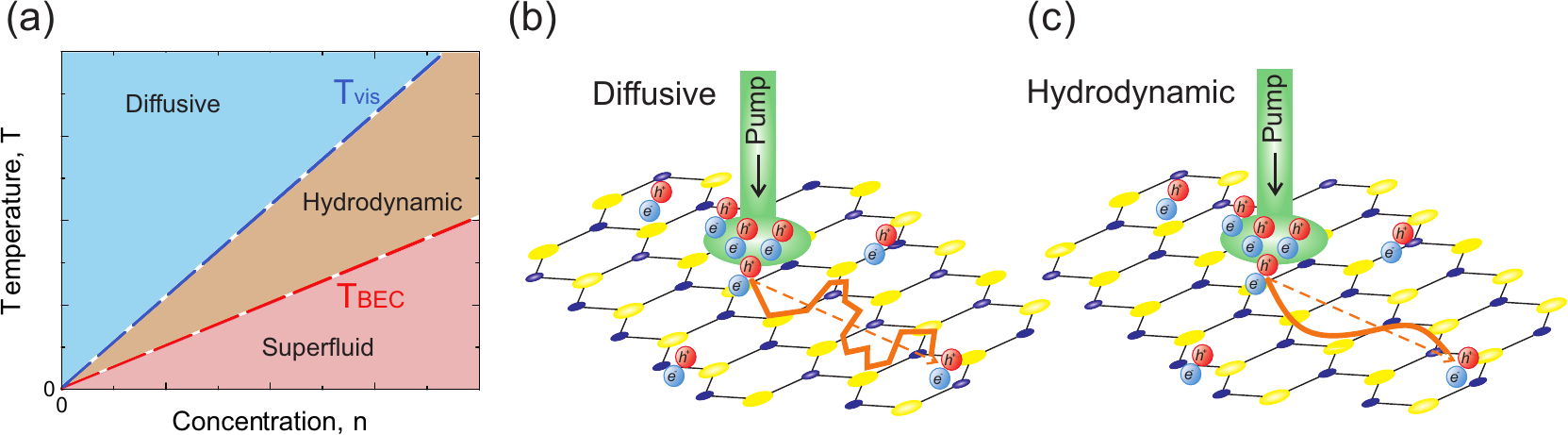}
\caption{\textbf{Exciton propagation regimes.} (a) Schematic phase diagram demonstrating an area of system parameters, where the viscous hydrodynamic regime takes place. Lines $T_{\textrm{vis}}(n)$ and $T_{\textrm{BEC}}(n)$ are plotted after Eqs.~\eqref{range:vis} and \eqref{range:BEC}, respectively. Sketches of (b) diffusive and (c) hydrodynamic regimes of exciton transport. The wiggles of trajectory (red curve) are shown not to scale with respect to the lattice constant, in reality the mean free path (wiggle size) is much larger than the lattice constant.}
\label{fig:scheme}
\end{figure*}

\section{Exciton propagation regimes and viscous hydrodynamics}\label{sec:phase}

\subsection{General analysis}\label{subsec:gen}

Figure \ref{fig:scheme} depicts the schematic phase diagram demonstrating a range of temperatures and densities, where different regimes of exciton propagation take place, including, the hydrodynamic regime [panel (a)]. Panels (b) and (c) schematically show the difference between diffusive and hydrodynamic regimes of exciton transport.

Viscous hydrodynamic regime of exciton transport appears in the system when exciton-exciton scattering is more efficient, than the exciton-phonon and exciton-defect scattering,
i.e., where
\begin{equation}
\label{hydro:crit}
\tau_{xx} \ll \tau_p.
\end{equation}
Furthermore, the criteria~\eqref{assumptions} should be fulfilled. In the state-of-the-art  transition metal dichalcogenide mono- and bilayers the main scattering mechanism at low temperatures ($2~\mbox{K} \lesssim T \lesssim 30~\mbox{K}$ is related to the exciton interaction with the long-wavelength acoustic phonons~\cite{Chernikov:2023ab,PhysRevLett.127.076801,Wietek2024}. The exciton momentum relaxation time by acoustic phonons with linear dispersion at not too low temperatures ($k_{B}T\gg Ms^{2}$) can be presented as \cite{Kaasbjerg2012,Shree2018,Glazov2020}
\begin{eqnarray} 
\label{tau:p:phonons}
\tau_{p}^{ac}=\frac{ms^{2}}{k_{B}T}\tau_{0}= \frac{c}{\hbar} T,
\end{eqnarray}
where $m$ is the exciton translational mass, $s$ is the speed of sound, $\tau_0$ is a temperature-independent constant related to the strength of the exciton-phonon interaction, and $c$ is a coefficient introduced in Ref.~\cite{Shree2018}. Naturally, $\tau_p^{ac} \propto 1/T$ because the larger the temperature the higher is the number of available phonons and $\tau_p^{ac}$ does not depend on the exciton density. By contrast, the exciton-exciton scattering time $\tau_{xx}$ derived in Eq.~\eqref{eq:ex_ex_scattering} is temperature independent and inversely proportional to the exciton density. Hence, the condition~\eqref{hydro:crit} provides on the exciton phase diagram in the coordinates $(n,T)$ the range 
\begin{equation}
\label{range:vis}
T<T_{{\textrm{vis}}} = \frac{n mU_0^2}{k_B \hbar^2} \frac{ms^2\tau_0}{\hbar},
\end{equation}
where the exciton-exciton interaction dominate over the exciton-phonon interactions. 
The dimensionless dependence $T_{{\textrm{vis}}}(n)$ is shown in the exciton phase diagram in Fig.~\ref{fig:scheme}(a) by dashed blue line. Further reduction of temperature makes exciton gas degenerate and Eq.~\eqref{non:deg} becomes violated. In that regime exciton ensemble turns to a collective state widely termed as Bose-Einstein condensate (BEC) with well known reservations in 2D systems, see, e.g.,~\cite{GlazovSuris_2021}. In any case, for sufficiently low temperatures~\cite{Fogler2014,PhysRevLett.105.070401} 
\begin{equation}
\label{range:BEC}
T<T_{{\textrm{BEC}}} \approx 1.3 \frac{\hbar^2 n}{k_B m} ,
\end{equation} 
the excitons should undergo Berezinskii-Kosterlitz-Thouless (BKT) transition and become superfluid.\footnote{Numerical coefficient in Eq.~\eqref{range:BEC} depends on details of exciton-exciton interactions, but this dependence is not dramatic.} The dependence $T_{{\textrm{BEC}}}(n)$ is plotted in Fig.~\ref{fig:scheme}(a) by  dashed red line. 

Importantly, both $T_{{\textrm{vis}}}$ and $T_{{\textrm{BEC}}}$ are proportionally to the same first power of the exciton density $n$. Their ratio is controlled by a dimensionless coefficient 
\begin{equation}
\label{dim:C}
\mathcal C = \frac{T_{{\textrm{BEC}}}}{T_{{\textrm{vis}}}}.
\end{equation}
For $\mathcal C>1$ the BEC/BKT phase is realized at higher temperatures than that of viscous hydrodynamics. In that case excitons turn from diffusive to superfluid with the decrease of the temperature at a fixed density or with the increase of the density at a fixed temperature. There is no viscous hydrodynamic range in the phase diagram in Fig.~\ref{fig:scheme}(a). By contrast, for $\mathcal C<1$ there is a range of temperatures where, at a fixed $n$, excitons form a viscous fluid. It is the situation of interest for us.

To summarize, for $\mathcal C<1$ there are three regimes of exciton propagation:
\begin{itemize}
\item{\emph{Diffusive regime}} is realized at sufficiently high temperatures and low densities where $T>T_{{\textrm{vis}}}(n)$;
\item{\emph{Viscous hydrodynamic regime}} is realized in the intermediate temperature and densitites range $T_{{\textrm{BEC}}}(n) <T< T_{{vis}}(n)$;
\item{\emph{Superfluid regime}} is realized at the low temperatures/high densities $T<T_{{\textrm{BEC}}}(n)$.
\end{itemize}

 We now analyze the possibility to observe the hydrodynamic regime of exciton transport in a single- and few-layer transition metal dichalcogenides systems and show that the case $\mathcal C<1$ can indeed be realized in such systems.

\begin{table*}[htb]
\caption{Relevant parameters of transition metal dichaclogenides (speed of sound $s$, the strength of exciton-phonon interaction $c$~Eq.~\eqref{tau:p:phonons}, translational mass $m$, ground state exciton binding energy $E_{1}$, exciton radius $R_{1}$) used in calculations, exciton-exciton interaction matrix element $U_{0}$ and ratio $\mathcal C = T_{{\textrm{BEC}}}/T_{{\textrm{vis}}}$ for hBN encapsulated monolayers calculated after Eqs.~\eqref{range:vis}, \eqref{range:BEC}, and \eqref{dim:C}.} \label{tab:param:mono}
\begin{ruledtabular}
\begin{tabular}{c|c|c|c|c|c|c|c}
Material/parameter& s, cm/sec \cite{Shree2018} & c, $\mu$eV/K \cite{Shree2018} & $m/m_0$ \cite{Han2018,Kormanyos2015} & $E_{1}$, meV \cite{Goryca2019} & $R_{1}$, nm \cite{Goryca2019} & $U_0$, meV$\cdot$nm$^{2}$ & $\mathcal C$\\
 \hline
$MoS_{2}$ & $6.6\times10^{5}$ & 45 & $0.99$ & 221 & 1.2 & $6.6\times10^{2}$ & 0.012\\
$MoSe_{2}$  & $4.1\times10^{5}$ & 52 & $1.13$ & 231 & 1.1 & $5.8\times10^{2}$ & 0.032\\
$WS_{2}$ & $4.3\times10^{5}$  & 60 & $0.62$ & 180 & 1.8 & $1.2\times10^{3}$ & 0.047\\
$WSe_{2}$ & $3.3\times10^{5}$ & 28 & $0.70$ & 167 & 1.7 & $9.9\times10^{2}$ & 0.038
\end{tabular}
\end{ruledtabular}
\end{table*}

\subsection{Monolayers of transition metal dichalcogenides}\label{subsec:mono}

In monolayer semiconductors similarly to single narrow quantum well structures the main contribution to exciton-exciton interaction results from the exchange interaction between identical carriers, i.e., from electron-electron and hole-hole exchange. The interaction is effective on the lengthscale on the order of exciton Bohr radius and effectively short-range, justifying Eq.~\eqref{V:inter}. The exciton-exciton interaction matrix element $U_{0}$ was analyzed in detail in \cite{Shahnazaryan2017} and the following expression was obtained
\begin{eqnarray} 
U_{0}=\alpha E_{n}R_{n}^{2},
\label{eq:matrix_element}
\end{eqnarray}
with $R_{n}$ and $E_{n}$ being the radius and energy of $n$-th exciton state, $\alpha=2.07$ is a coefficient. In what follows we consider only $1s$ ground exciton states. Using the material parameters from Refs. \cite{Shree2018} and \cite{Goryca2019} we calculate the matrix element $U_0$ for four widespread transition metal dichalcogenide monolayers and, subsequently, the ratio $\mathcal C$ in Eq.~\eqref{dim:C}, see Tab.~\ref{tab:param:mono}. It follows from our analysis that $\mathcal C<1$ for all studied systems. Therefore, there is a range of exciton densities where viscous hydrodynamic regime can be realized before system becomes degenerate and superfluid.

To give an example, for $n =1\times 10^{11}$~cm$^{-2}$ in $WS_2$ monolayer and $T=4$~K the exciton-phonon scattering time [Eq.~\eqref{tau:p:phonons}] is $\tau_p = 5.3\times10^{-13}$~s and exciton-exciton scattering time [Eq.~\eqref{eq:ex_ex_scattering}] is $\tau_{xx} = 5.6\times10^{-14}$~s being much shorter than $\tau_p$. Hence, the excitons indeed become a viscous fluid for reasonable parameters in transition metal dichalcogenide monolayers. For this density the excitons are non-degenerate and far below the Mott transition.

\subsection{Bi-layers of transition metal dichalcogenides}\label{subsec:bi}

In bilayer systems the exciton-exciton interactions are expected to be stronger~\cite{GlazovSuris_2021} and, hence, the condition for dominant exciton-exciton scattering and, accordingly, viscous hydrodynamics, $\mathcal C<1$, should be better pronounced for such systems. The key contribution to the exciton-exciton interaction in heterobilayers case is expected to be provided by the dipole-dipole repulsion of excitons that can be roughly evaluated as
\begin{equation}
\label{dd:rep}
V(\mathbf r - \mathbf r') \approx \frac{e^2 d^2}{\varkappa |\mathbf r- \mathbf r'|^3}, \quad |\mathbf r- \mathbf r'|\gtrsim d,
\end{equation}
where $d$ is the effective interlayer distance and $\varkappa$ is the effective dielectric constant.
Despite a power-law decay of the exciton-exciton interaction potential the integral $\int V(\mathbf r) d\mathbf r$ converges at $r\to \infty$ making it possible to model the dipole-dipole interaction as a short-range, cf. Eq.~\eqref{V:inter}. Equation~\eqref{dd:rep}, however, strongly overestimates exciton-exciton repulsion owing to the fact that in heterobilayers $d$ and exciton Bohr radius are of the same order of magnitude and because of manybody effects~\cite{GlazovSuris_2021,Wietek2024,Erkensten2021,Steinhoff2023}.

Thus, it is natural to estimate $U_0$ from the experimental data.  For $MoSe_{2}-WSe_{2}$ bilayers $U_0 \sim 10^{2}$~meV$\cdot$nm$^{2}$ was evaluated in Ref.~\cite{Wietek2024}. Obtained value is rather small, in the same ballpark of values as for the monolayers, Tab.~\ref{tab:param:mono}, but it can be enhanced by increasing the spacing between the layers or applying electric field perpendicular to the bilayer \cite{Tagarelli2023,Sun2022}. For example, application of vertical electric field $E_{z}=300$ mV$\cdot$nm$^{-1}$ to the $WSe_{2}$ homobilayer leads to the increase of constant $U_0$ up to
$1.2\times10^{3}$ meV$\cdot$nm$^{2}$. This value is very close to some of the values reported earlier, such as  $1.6\times10^{3}$ meV$\cdot$nm$^{2}$ \cite{Yuan2020} and
$2.6\times10^{3}$ meV$\cdot$nm$^{2}$ \cite{Sun2022} in $WS_{2} - WSe_{2}$ and $WS_{2} - MoSe_{2}$ heterobilayers, correspondingly. All-in-all, for heterobilayers, in particular, in the presence of electric field normal to the layers, the parameter $\mathcal C\ll 1$ demonstrating a wide the range of viscous hydrodynamics in heterobilayers.

\section{Exciton propagation: results}\label{sec:transport}

The derived exciton propagation equations~\eqref{excitons:hydro} encompass all relevant regimes of exciton propagation in 2D semiconductor systems: diffusive, viscous hydrodynamic, and superfluid. We briefly address these regimes one after the other.

In the case of \emph{diffusive propagation} of excitons the momentum scattering processes dominate over the exciton-exciton scattering and Eq.~\eqref{hydro:crit} is violated. In the diffusive regime $\tau_p \ll \tau_{xx}$. It allows us to omit in the right-hand side of Eq.~\eqref{flux:fin} (or of Eq.~\eqref{NS}) the contributions to the exciton flux density tensor related to the viscosity and account for the pressure term only. For the same reasons the term with the temporal derivative $\partial (n\mathbf V)/\partial t$ can be disregarded. It brings us to the standard drift-diffusion picture of exciton propagation~\cite{Chernikov:2023ab,Wietek2024} with
\begin{equation}
\label{Ficks}
\mathbf V = -\frac{k_B T\tau_p}{m n} \frac{\partial n}{\partial \mathbf r} - \frac{\tau_p}{m} \frac{\partial U}{\partial \mathbf r}.
\end{equation}
Equation~\eqref{Ficks} is nothing but Fick's law of diffusion that provides the diffusion coefficient $D= k_B T \tau_p/m$ and the effective exciton mobility $e\tau_p/m$ (with $e$ being elementary charge) in agreement with Einstein's relation. The validity of the diffusive approach is justified in the case where the spatial and temporal gradients of the exciton distribution function are small
\[
\left|\frac{1}{n(\mathbf r,t)} \frac{\partial n(\mathbf r,t)}{\partial \mathbf r}\right| \ll \sqrt{\frac{k_B T}{m}}\tau_p, 
\]
\[
\left|\frac{1}{n(\mathbf r,t)} \frac{\partial n(\mathbf r,t)}{\partial t}\right| \ll \sqrt{\frac{k_B T}{m}}\tau_p.
\]
It follows from combining Eqs.~\eqref{cont:fin} and Eq.~\eqref{Ficks} the generalized drift-diffusion equation~\cite{Chernikov:2023ab,Wietek2024}
\begin{eqnarray}
\frac{\partial n}{\partial t}=D\Delta n-\frac{n}{\tau}-R_{A}n^{2}+\frac{D}{k_{B}T}\nabla\cdot(n\nabla U),
\label{eq:diffusion}
\end{eqnarray}
where we took into account both mono- and bimolecular recombination of excitons described by the time constants $\tau$ and $R_A$, respectively~\cite{Kulig2018}.

In the \emph{superfluid} regime both the momentum scattering and exciton-exciton collisions are negligible. In that case the viscous stress vanishes, $\sigma_{\alpha\beta} =0$, and Eq.~\eqref{NS} where one has to put $\eta=0$, $\tau_p \to \infty$ reduces to the Euler equation describing a flow of an ideal liquid, cf.~\cite{GlazovSuris_2021}.

\begin{figure}[t]
  \includegraphics[width=1.0\linewidth]{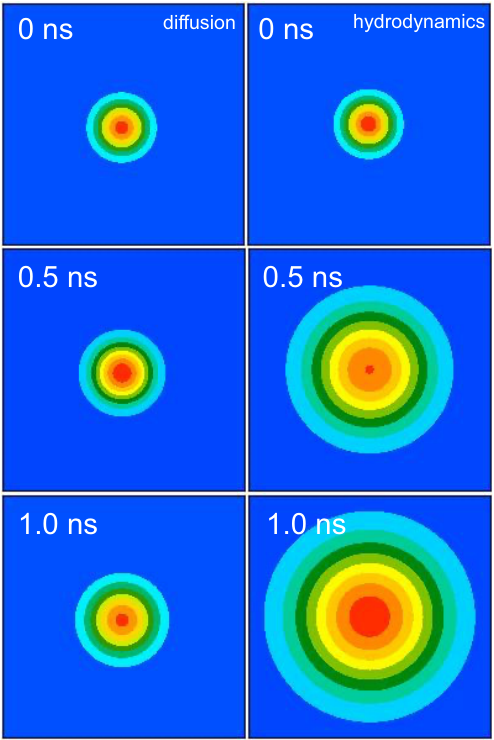}
  \caption{\textbf{Exciton cloud expansion.} Panels (a) --- (c) in the left column demonstrate time evolution of exciton density in the diffusive regime obtained by numerical solution of Eq~\eqref{eq:diffusion}. Panels (d) --- (f) in the right column show time evolution of exciton density in the hydrodynamic regime from numerical solution of Eqs.~\eqref{excitons:hydro}. Snapshots correspond to the time moments $t= 0$ ns, 0.5 ns, 1.0 ns. Color bars correspond to the concentration of excitons in the units of cm$^{-2}$. The size of each panel is $7\times 7$ $\mu$m$^{2}$. Modelling is performed for initial concentration of excitons $1.0\times10^{11}$~cm$^{-2}$, the excitation spot size $r_0=0.4$ $\mu$m, $\tau/\tau_{p}=20$, and $m=0.62m_{e}$. Parameters for panels (a)-(c) are the following: linear diffusion coefficient $D=3$ cm$^{2}/$s, Auger recombination rate $R_{A}=10^{-1}$~cm$^{2}/$s, and effective repulsion constant $U_0D/{k_{B}T}=10^{-10}$ cm$^{4}/$sec, recombination time $\tau=0.5$~ns. Parameters for panels (d)-(f) are: exciton interaction matrix element $U_{0}=1.2\times10^{3}$~meV$\cdot$nm$^{2}$, $T=4$ K.}
  \label{fig:diffusion_hydrodynamics}
\end{figure}

Between these two regimes there is a \emph{viscous hydrodynamic} regime of exciton propagation where excitons propagate as a fluid, but the dissipation is caused  by the momentum scattering and viscosity. The latter provides the transfer of momentum from the center of exciton cloud to the periphery where the scattering can occur, e.g., at the channel edges. Moreover, exciton-exciton repulsion enhances the effective propagation speed of excitons in this regime, as illustrated in Fig.~\ref{fig:diffusion_hydrodynamics} where the three panels (a) --- (c) in the left column show the exciton expansion in the diffusive regime while the three panels (d) --- (f) in the right column show the exciton expansion in the viscous hydrodynamic regime. A clear increase in the expansion rate in the viscous hydrodynamic regime is seen.

To have a qualitative and semi-quantitative description of exciton propagation in all free studied regimes we derive the equation for the characteristic radius of the exciton cloud $R(t)$ defined as [cf. Refs. \cite{Wietek2024,Kuznetsov:2020aa} and references therein]
\begin{eqnarray}
[R(t)]^2 \equiv R_2(t) =\frac{1}{N}\int r^{2}n(\textbf{r},t)d\textbf{r},
\end{eqnarray}
where $N=\int n(\mathbf{r})d\textbf{r}$ is the total number of excitons. Neglecting the exciton decay processes we, first, multiply the continuity Eq.~\eqref{cont:fin} by $d/d t + \tau_p^{-1}$ and, second, multiply the derived equation by $r^2$ and, third, integrate it over the sample area. As a result, we have~\cite{Wietek2024}
\begin{eqnarray}
\frac{d}{d t}\left(\frac{d}{d t}+\frac{1}{\tau_p}\right)R_{2}(t)=\frac{4\mathcal E_{\rm tot}(t)}{mN},
\label{eq:hydro_eq}
\end{eqnarray}
where 
\begin{equation}
    \label{Etot}
    \mathcal E_{\rm tot} = m\int P d\mathbf{r}.
\end{equation}
is the total energy of the exciton gas.

\begin{figure}[b]
  \includegraphics[width=0.95\linewidth]{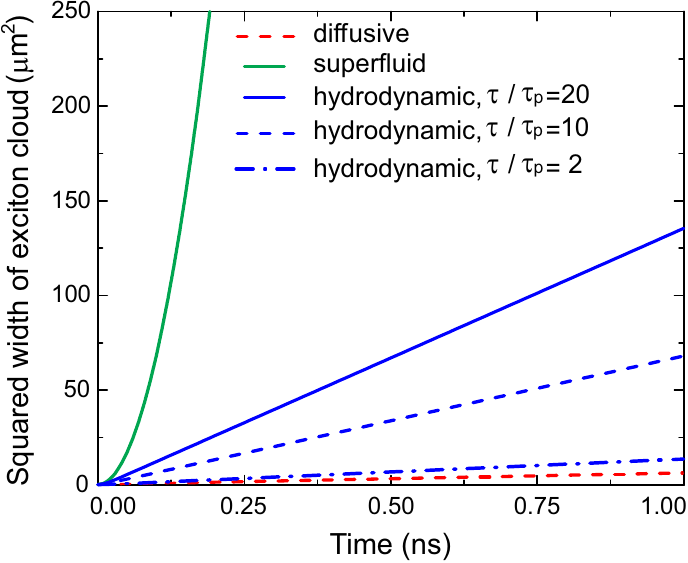}
  \caption{\textbf{Exciton cloud area.} Squared size of exciton cloud, $R^2(t)$, as a function of time calculated after Eq.~\eqref{eq:hydro_R}. Solid green curve shows superfluid regime,  dashed red curve corresponds to the diffusion regime and blue curves demonstrate hydrodynamic regime. Solid blue curve is obtained for $\tau/\tau_p=20$, dashed blue line is obtained for $\tau/\tau_p=10$ and dashed-dotted blue line is obtained for $\tau/\tau_{p}=2$. For all the curves the material parameters correspond to $WS_{2}$: $m=0.62m_e$, and $U_{0}=1.2\times10^{3}$ meV$\cdot$nm$^{2}$. Initial exciton density  $n=1.0\times10^{11}$ cm$^{-2}$, excitation spot size $r_0=0.4$~$\mu$m. }
  \label{fig:excitons_cloud}
\end{figure}

\begin{figure*}[tb]
  \includegraphics[width=0.8\linewidth]{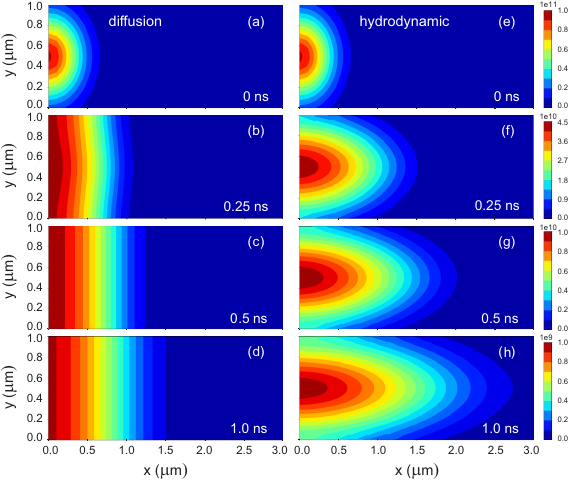}
  \caption{\textbf{Exciton expansion in 2D channel.} Panels (a) --- (c) in the left column demonstrate time evolution of exciton density in the diffusive regime obtained by numerical solution of Eq~\eqref{eq:diffusion}. Panels (d) --- (f) in the right column show time evolution of exciton density in the hydrodynamic regime from numerical solution of Eqs.~\eqref{excitons:hydro}. Snapshots correspond to the time moments $t= 0$ ns, 0.5 ns, 1.0 ns. Color bars correspond to the concentration of excitons in the units of cm$^{-2}$. The size of each panel is $0.5\times 3.0$ $\mu$m$^{2}$. Modelling is performed for initial concentration of excitons $1.0\times10^{11}$~cm$^{-2}$ and excitation spot size $r_0=0.4$ $\mu$m, $\tau/\tau_{p}=20$, and $m=0.62m_{e}$. Parameters for panels (a) --- (c) are the following: linear diffusion coefficient $D=3$ cm$^{2}/$s, Auger recombination rate $R_{A}=10^{-1}$~cm$^{2}/$s, and effective repulsion constant $U_0D/{k_{B}T}=10^{-10}$ cm$^{4}/$s, recombination time $\tau=0.5$~ns. Parameters for panels (d) --- (f) are: exciton interaction matrix element $U_{0}=1.2\times10^{3}$ meV$\cdot$nm$^{2}$, $T=4$~K.}
  \label{fig:channel}
\end{figure*}

If the excitons were completely isolated from environment, the total energy is conserved and Eq.~\eqref{eq:hydro_eq} can be readily integrated under conditions $R(0)=r_0$, $dR/dt|_{t=0}=0$ with the result
\begin{multline}
R(t) = \sqrt{R_2(t)} \\
= \left[r_0^2 +  \frac{4\tau_p^2 \mathcal E_{tot}}{mN}\left(t/\tau_p + e^{-t/\tau_p} -1\right)\right]^{1/2}.
\label{eq:hydro_R}
\end{multline}
At long times $t\gg \tau_p$ the spread of excitons is diffusive: It follows from Eq.~\eqref{eq:hydro_R} that $R(t) = 2\sqrt{D_{\rm eff} t}$ with the effective diffusion coefficient determined by the total energy $D_{\rm eff} = \mathcal E_{\rm tot}\tau_p/(mN)$. For negligible interactions $D_{\rm eff} \to D$. At short times (and also in the superfluid case where $\tau_p\to \infty$) the expansion occurs with a constant velocity $R(t) \approx  v_{\rm eff} t$ where $v_{\rm eff} = \sqrt{2\mathcal E_{\rm tot}/(mN)}$.

Temporal dependence of the squared exciton cloud size, $R^2(t)$, found from Eq.~\eqref{eq:hydro_R} is illustrated in Fig.~\ref{fig:excitons_cloud}. It is clearly evident that spreading of exciton cloud in hydrodynamic regime is faster, than in the diffusive regime, but much slower, than in the superfluid regime, where both viscosity and momentum relaxation processes are neglected.

While quantiative variations of the exciton propagation speed between the diffusive and hydrodynamic regimes are, as seen from Fig.~\ref{fig:excitons_cloud},  striking, qualitatively, the shape of exciton cloud remains basically the same, see Fig.~\ref{fig:diffusion_hydrodynamics}. It is instructive to consider another possible future experimental setting where the difference between the diffusive and viscous hydrodynamic propagation is qualitative. Such situation can be realized in a narrow channel whose width is comparable with or smaller than the exciton mean free path. The illustration of exciton propagation in this case is presented in Fig.~\ref{fig:channel}. Compared to a flat front of the exciton density in the case of diffusive propagation, in the viscous hydrodynamics regime, the front curves to a parabola-like shape reflecting the parabolic profile of the Poiseuille flow of viscous liquid. While this regime is hard to realize in the state-of-the-art samples, a careful engineering of the structure and selection of parameters may, in future, enable one to realize such a regime.

\section{Conclusion}
\label{sec:conclusion}
We have theoretically demonstrated the possibility to reach the hydrodynamic regime of exciton propagation in two-dimensional semiconductors based on transition metal dichalcogenides mono- and bilayers. Hydrodynamic-like equations for exciton propagation describing viscous flow of interacting excitons have been derived based on the Boltzmann kinetic equation for the exciton distribution function and the exciton liquid viscosity has been calculated. Under reasonable conditions, the exciton-exciton collisions could dominate over exciton-phonon and static disorder scattering in mono- and bi-layers of transition metal dichalcogenides making it possible to reach the viscous hydrodynamic regime of exciton propagation, where the exciton expansion occurs faster than in the diffusive case. We have compared various exciton
propagation regimes and established the conditions, which allow one to observe the hydrodynamic regime of exciton transport.

\section{Acknowledgements}

 V.N.M. thanks for support Russian Science Foundation grant No. 24-12-00020.

\newpage
\begin{widetext}
\appendix

\section{Second angular harmonic relaxation at exciton-exciton collisions}\label{app:xx}

Let us consider non-equilibrium occupation of exciton states with anisotropic distribution function related to the velocity gradient in the form [cf. Eqs.~\eqref{sigma:0}, \eqref{sigma} and \eqref{dfk:temp}]
\begin{eqnarray}
\label{eq:app:df}
f_{\mathbf{k}}=f_{\mathbf{k}}^{0}+Bk_{x}k_{y}f_{\mathbf{k}}^{0}=f_{\mathbf{k}}^{0}(1+Bk_{x}k_{y})
\end{eqnarray}
with Boltzmann equilibrium distribution in the form $f_{\mathbf{k}}^{0}=\exp[({\mu-\varepsilon_{k}})/{k_{B}T}]$. In this case collision integral reads
\begin{eqnarray}
Q_{xx}\{v_{\mathbf{k}}\}=-\frac{4\pi|U_{0}|^{2}}{S^{2}\hbar}\sum_{\mathbf{k}'\mathbf{p}\mathbf{p}'}\delta_{\mathbf{k}+\mathbf{p},\mathbf{k}'+\mathbf{p}'}\delta(\varepsilon_{k}+\varepsilon_{p}-\varepsilon_{k'}-\varepsilon_{p'})\times(v_{\mathbf{k}}f_{\mathbf{p}}-v_{\mathbf{k}'}f_{\mathbf{p}'}),
\end{eqnarray}
where $v_{\mathbf{k}}=f_{\mathbf{k}}^{0}(1+Bk_{x}k_{y})$ and $S$ is the normalization area, which we set to be equal to unity. In order to determine the stress tensor, exciton collision time and exciton viscosity we need to find the projection on the second angular harmonics as $\sin(2\phi_{\mathbf{k}})=2\sin\phi_{\mathbf{k}}\cos\phi_{\mathbf{k}}=2k_{x}k_{y}/k^{2}$.
So, 

\begin{eqnarray}
\label{tau:xx:A3}
\frac{1}{\tau_{xx}}=\frac{\sum_{\mathbf{k}}\sin(2\phi_{\mathbf{k}})Q_{xx}\{k_{x}k_{y}f_{\mathbf{k}}^{0}\}}{\sum_{\mathbf{k}}\sin(2\phi_{\mathbf{k}})k_{x}k_{y}f_{\mathbf{k}}^{0}},
\end{eqnarray}
where we have disregarded the terms $\propto f_{\mathbf k}^0$ since the collision integral vanishes at the equilibrium distribution function. Performing calculations we follow the procedure discussed in details in Refs.~\cite{ll10_eng} and \cite{PhysRevB.106.235305} in calculation of the viscosity.
The sum in the numerator of Eq.~\eqref{tau:xx:A3} is evaluated as:
\begin{eqnarray}
&&\sum_{\mathbf{k}}\sin(2\phi_{\mathbf{k}})Q_{xx}\{k_{x}k_{y}f_{\mathbf{k}}^{0}\}=-\frac{4\pi|U_{0}|^{2}}{\hbar}\sum_{\mathbf{k}'\mathbf{p}\mathbf{p}'}\delta_{\mathbf{k}+\mathbf{p},\mathbf{k}'+\mathbf{p}'}\delta(\varepsilon_{k}+\varepsilon_{p}-\varepsilon_{k'}-\varepsilon_{p'})\nonumber\\&&
\times\frac{2k_{x}k_{y}}{k^{2}}\cdot[(1+Bk_{x}k_{y})f_{\mathbf{k}}^{0}(1+Bp_{x}p_{y})f_{\mathbf{p}}^{0}-(1+Bk'_{x}k'_{y})f_{\mathbf{k'}}^{0}(1+Bp'_{x}p'_{y})f_{\mathbf{p'}}^{0}]=/\mathbf{k(p)}\leftrightarrow \mathbf{k'(p')}/\nonumber\\&&=-\frac{4\pi|U_{0}|^{2}}{\hbar}\sum_{\mathbf{k}'\mathbf{p}\mathbf{p}'}\delta_{\mathbf{k}+\mathbf{p},\mathbf{k}'+\mathbf{p}'}\delta(\varepsilon_{k}+\varepsilon_{p}-\varepsilon_{k'}-\varepsilon_{p'})\cdot\biggl[\frac{2k_{x}k_{y}}{k^{2}}-\frac{2k'_{x}k'_{y}}{k'^{2}}\biggr]\nonumber\\&&\times[1+B(k_{x}k_{y}+p_{x}p_{y})+B^{2}k_{x}k_{y}p_{x}p_{y}]f_{\mathbf{k}}^{0}f_{\mathbf{p}}^{0}=
-\frac{4\pi|U_{0}|^{2}}{\hbar}\cdot[X_{2a}-X'_{2a}+B(X_{2b}-X'_{2b})+B^{2}(X_{2c}-X'_{2c})].\nonumber\\
\label{app:sum}
\end{eqnarray}
Further, we'll calculate the integrals contributing to Eq.(\ref{app:sum}). Terms $X_{2c}$ and $X'_{2c}$ should be omittted as their contribution is $O(B^{2})$. Taking into account the following relations:

\begin{eqnarray}
\frac{1}{2\pi}\int_{0}^{2\pi}d\theta e^{z \cos\theta}\cos(n\theta)&&=I_{n}(z),\nonumber\\
\int_{0}^{\infty}I_{n}(at)e^{-p^{2}t^{2}}t^{n+1}dt&&=\frac{a^{n}\exp(\frac{a^{2}}{4p^{2}})}{(2p^{2})^{n+1}}
\end{eqnarray}
and introducing $k_{T}^{2}=\frac{2mk_{B}T}{\hbar^{2}}$ with $m$ being a translational mass of exciton, we analyze the main contribution to exciton-exciton scattering given by the terms $X_{2b}$ and $X'_{2b}$ in Eq.(\ref{app:sum}). 
Introducing
\begin{eqnarray}
&&k_{x}=k\cos\phi_{\mathbf{k}},\nonumber\\
&&k_{x}\varkappa_{y}=k\varkappa \cos\phi_{\mathbf{k}}\sin(\phi_{\mathbf{k}}+\theta),\nonumber\\
&&\varkappa_{x}\varkappa_{y}=\varkappa^{2}\cos(\phi_{\mathbf{k}}+\theta)\sin(\phi_{\mathbf{k}}+\theta),
\end{eqnarray}
one can get
\begin{eqnarray}
&&X_{2b}=\sum_{\mathbf{k}\mathbf{k}'\mathbf{p}\mathbf{p}'}\delta_{\mathbf{k}+\mathbf{p},\mathbf{k}'+\mathbf{p}'}\delta(\varepsilon_{k}+\varepsilon_{p}-\varepsilon_{k'}-\varepsilon_{p'})\cdot\frac{2k_{x}k_{y}(k_{x}k_{y}+p_{x}p_{y})}{k^{2}}f_{\mathbf{k}}^{0}f_{\mathbf{p}}^{0}\nonumber\\
&&=\frac{m}{\hbar^{2}}\sum_{\mathbf{k},\mathbf{\varkappa},\mathbf{\varkappa}'}\delta(\varkappa^{2}-\varkappa'^{2})\cdot\frac{2k_{x}k_{y}}{k^{2}}\times\frac{2k_{x}k_{y}+\varkappa_{x}\varkappa_{y}-k_{x}\varkappa_{y}-k_{y}\varkappa_{x}}{k^{2}}f_{\mathbf{k}}^{0}f_{\mathbf{k}-\mathbf{\varkappa}}^{0}=/\theta=\angle(\mathbf{k},\mathbf{\varkappa})/\nonumber\\&&
=\frac{\pi m}{\hbar^{2}}\frac{1}{(2\pi)^{4}}\exp\left(\frac{2\mu}{k_{B}T}\right)\sum_{\mathbf{k}}\frac{2k_{x}k_{y}}{k^{2}}\exp\left(-\frac{2k^{2}}{k_{T}^{2}}\right)\int_{0}^{\infty}\varkappa d\varkappa \exp\left(-\frac{\varkappa^{2}}{k_{T}^{2}}\right)\int_{0}^{2\pi}d \theta \exp\left(\frac{2k \varkappa \cos\theta}{k_{T}^{2}}\right)\nonumber\\&&\times
[k^{2}\sin(2\phi_{\mathbf{k}})+\frac{\varkappa^{2}}{2}\sin(2\phi_{\mathbf{k}})\cos(2\theta)+\frac{\varkappa^{2}}{2}\cos(2\phi_{\mathbf{k}})\sin(2\theta)-k\varkappa \sin(2\phi_{\mathbf{k}}\cos\theta-k\varkappa \cos(2\phi_{\mathbf{k}}\sin\theta)]\nonumber\\&&=\frac{\pi^{2} m}{\hbar^{2}} \frac{1}{(2\pi)^{6}}\exp\left(\frac{2\mu}{k_{B}T}\right)\int_{0}^{\infty}kdk\exp\left(-\frac{2k^{2}}{k_{T}^{2}}\right)\int_{0}^{2\pi}\sin^{2}(2\phi_{\mathbf{k}})d\phi_{\mathbf{k}}\int_{0}^{\infty}\varkappa d\varkappa \exp\left(-\frac{\varkappa^{2}}{k_{T}^{2}}\right)\nonumber\\&&\times[k^{2}I_{0}\left(\frac{2k\varkappa}{k_{T}^{2}}\right)-k\varkappa I_{1}\left(\frac{2k\varkappa}{k_{T}^{2}}\right)+\frac{\varkappa^{2}}{2}I_{2}\left(\frac{2k\varkappa}{k_{T}^{2}}\right)]
=/x=k/k_{T}, y=\varkappa/k_{T}/\nonumber\\&&=\frac{2\pi^{3}mk_{T}^{6}}{\hbar^{2}}\frac{1}{(2\pi)^{6}}\exp\left(\frac{2\mu}{k_{B}T}\right)\int_{0}^{\infty}xdxe^{-2x^{2}}\int_{0}^{\infty}ydye^{-y^{2}}\cdot[x^{2}I_{0}\left(2xy\right)-xyI_{1}\left(2xy\right)+\frac{y^{2}}{2}I_{2}\left(2xy\right)]\nonumber\\&&=
\frac{2\pi^{3}mk_{T}^{6}}{\hbar^{2}}\frac{1}{(2\pi)^{6}}\exp\left(\frac{2\mu}{k_{B}T}\right)\int_{0}^{\infty}xdxe^{-2x^{2}}\cdot[x^{2}\frac{e^{x^{2}}}{2}-x\frac{xe^{x^{2}}}{2}+\frac{1}{2}\frac{x^{2}e^{x^{2}}}{2}]=\frac{\pi^{3}mk_{T}^{6}}{4\hbar^{2}} \frac{1}{(2\pi)^{6}}\exp\left(\frac{2\mu}{k_{B}T}\right)
\end{eqnarray}
One can calculate $X'_{2b}$ in the similar way considering
\begin{eqnarray}
&&k_{x,}=k,\cos\phi_{\mathbf{k},},\nonumber\\
&&\varkappa_{x}=\varkappa \cos\phi_{\mathbf{\varkappa}},\nonumber\\
&&\varkappa'_{x}=\varkappa \cos(\phi_{\mathbf{k}'}+\theta),
\end{eqnarray}
one can get
\begin{eqnarray}
&&X'_{2b}=\sum_{\mathbf{k}\mathbf{k}'\mathbf{p}\mathbf{p}'}\delta_{\mathbf{k}+\mathbf{p},\mathbf{k}'+\mathbf{p}'}\delta(\varepsilon_{k}+\varepsilon_{p}-\varepsilon_{k'}-\varepsilon_{p'})\cdot\frac{2k'_{x}k'_{y}(k_{x}k_{y}+p_{x}p_{y})}{k'^{2}}f_{\mathbf{k}}^{0}f_{\mathbf{p}}^{0}\nonumber\\
&&=\frac{m}{\hbar^{2}}\sum_{\mathbf{k},\mathbf{\varkappa},\mathbf{\varkappa}'}\delta(\varkappa^{2}-\varkappa'^{2})\cdot\frac{2k'_{x}k'_{y}}{k'^{2}}\times\left[(k'_{x}+\frac{\varkappa_{x}-\varkappa'_{x}}{2})(k'_{y}+\frac{\varkappa_{y}-\varkappa'_{y}}{2})-(k'_{x}-\frac{\varkappa_{x}+\varkappa'_{x}}{2})(k'_{y}-\frac{\varkappa_{y}+\varkappa'_{y}}{2})\right]\nonumber\\
&&\times f_{\mathbf{k}'+\frac{1}{2}(\mathbf{\varkappa}-\mathbf{\varkappa}')}^{0}f_{\mathbf{k}'-\frac{1}{2}(\mathbf{\varkappa}+\mathbf{\varkappa}')}^{0}=/\theta=\angle(\mathbf{k}',\mathbf{\varkappa}')/=\frac{m}{\hbar^{2}}\frac{1}{(2\pi)^{4}}\exp\left(\frac{2\mu}{k_{B}T}\right)\sum_{\mathbf{k}'}\frac{2k'_{x}k'_{y}}{2k'^{2}}\exp\left(-\frac{2k'^{2}}{k_{T}^{2}}\right)
\int_{0}^{\infty}\varkappa d\varkappa\exp\left(-\frac{\varkappa^{2}}{k_{T}^{2}}\right)\nonumber\\&&\times\int_{0}^{2\pi}d\phi_{\mathbf{\varkappa}}\int_{0}^{2\pi}d\theta \exp\left(\frac{2k'\varkappa \cos\theta}{k_{T}^{2}}\right)\left[k'_{y}\varkappa_{x}+k'_{x}\varkappa_{y}-\frac{\varkappa\varkappa_{x}}{2}\sin(\phi_{\mathbf{k}'}+\theta)-\frac{\varkappa\varkappa_{y}}{2}\cos(\phi_{\mathbf{k}'}+\theta)\right]\nonumber\\&&=\frac{\pi m}{\hbar^{2}}\frac{1}{(2\pi)^{6}}\exp\left(\frac{2\mu}{k_{B}T}\right)\int_{0}^{\infty}k'dk'\exp\left(-\frac{2k'^{2}}{k_{T}^{2}}\right)\int_{0}^{\infty}\varkappa d\varkappa \exp\left(-\frac{\varkappa^{2}}{k_{T}^{2}}\right)\cdot\left[k\varkappa I_{0}\left(\frac{2k\varkappa}{k_{T}^{2}}\right)-\frac{\varkappa^{2}}{2}I_{1}\left(\frac{2k\varkappa}{k_{T}^{2}}\right)\right]\nonumber\\&&\times\int_{0}^{2\pi}d\phi_{\mathbf{k}'}\sin(2\phi_{\mathbf{k}'})\int_{0}^{2\pi}d\phi_{\mathbf{\varkappa}}\sin(\phi_{\mathbf{\varkappa}}+\phi_{\mathbf{k}'})=0
\end{eqnarray}
as $\int_{0}^{2\pi}d\phi_{\mathbf{\varkappa}}\sin(\phi_{\mathbf{\varkappa}}+\phi_{\mathbf{k}'})=0$.
Let us now calculate the denominator
\begin{eqnarray}
&&\sum_{\mathbf{k}}\sin(2\phi_{\mathbf{k}})k_{x}k_{y}f_{\mathbf{k}}^{0}=\frac{1}{(2\pi)^{2}}\exp\left(\frac{\mu}{k_{B}T}\right)\int_{0}^{\infty}kdk\int_{0}^{2\pi}d\phi_{\mathbf{k}}k^{2}\frac{\sin^{2}(2\phi_{\mathbf{k}})}{2}\exp\left(\frac{k^{2}}{k_{T}^{2}}\right)\nonumber\\&&
=/\frac{k^{2}}{k_{T}^{2}}=t/=\frac{\pi k_{T}^{4}}{4}\frac{1}{(2\pi)^{2}} \exp\left(\frac{\mu}{k_{B}T}\right)\int_{0}^{\infty}te^{-t}dt=\frac{\pi k_{T}^{4}}{4}\frac{1}{(2\pi)^{2}} \exp\left(\frac{\mu}{k_{B}T}\right).
\end{eqnarray}

The exciton-exciton scattering rate is determined by the following way:

\begin{eqnarray}
\frac{1}{\tau_{xx}}&&=\frac{\frac{4\pi|U_{0}|^{2}}{\hbar}X_{2b}}{\frac{\pi k_{T}^{4}}{4}\frac{1}{(2\pi)^{2}}}\exp\left(\frac{\mu}{k_{B}T}\right)=\frac{16|U_{0}|^{2}}{\hbar k_{T}^{4}}\frac{X_{2b}}{\frac{1}{(2\pi)^{2}}}\exp\left(-\frac{\mu}{k_{B}T}\right)=\frac{m^{2}k_{B}T|U_{0}|^{2}}{2\pi\hbar^{5}}\exp\left(-\frac{\mu}{k_{B}T}\right).
\end{eqnarray}
Considering $f_{\mathbf{k}}^{0}=\exp(\frac{\mu-\varepsilon_{k}}{k_{B}T})=\frac{n}{gk_{B}T}\exp\left(-\frac{\varepsilon_{k}}{k_{B}T}\right)$ one can get $\exp\left(-\frac{\varepsilon_{k}}{k_{B}T}\right)=\frac{n}{gk_{B}T}$, where $n$ is the
particles density and density of states can be written as $g=\frac{m}{2\pi\hbar^{2}}$. So, one can finally get an expression describing the exciton-exciton scattering rate in the following form

\begin{eqnarray}
\frac{1}{\tau_{xx}}&&=\frac{m^{2}k_{B}T|U_{0}|^{2}}{2\pi\hbar^{5}}\frac{n}{gk_{B}T}=\frac{m^{2}k_{B}T|U_{0}|^{2}}{2\pi\hbar^{5}}\frac{2\pi\hbar^{2}n}{mk_{B}T}=\frac{n m|U_{0}|^{2}}{\hbar^{3}}.
\end{eqnarray}

\section{Appendix B}

Let us demonstrate straightforwardly, that contributions from  $X_{2a}$ and $X'_{2a}$ are equal to zero.
\begin{eqnarray}
&&X_{2a}=\sum_{\mathbf{k}\mathbf{k}'\mathbf{p}\mathbf{p}'}\delta_{\mathbf{k}+\mathbf{p},\mathbf{k}'+\mathbf{p}'}\delta(\varepsilon_{k}+\varepsilon_{p}-\varepsilon_{k'}-\varepsilon_{p'})\cdot\frac{2k_{x}k_{y}}{k^{2}}f_{\mathbf{k}}^{0}f_{\mathbf{p}}^{0}=
\frac{m}{\hbar^{2}}\sum_{\mathbf{k},\mathbf{\varkappa},\mathbf{\varkappa}'}\delta(\varkappa^{2}-\varkappa'^{2})\cdot\frac{2k_{x}k_{y}}{k^{2}}f_{\mathbf{k}}^{0}f_{\mathbf{k}-\mathbf{\varkappa}}^{0}\nonumber\\&&=/\theta=\angle(\mathbf{k},\mathbf{\varkappa})/
=\frac{\pi m}{\hbar^{2}}\frac{1}{(2\pi)^{2}}\exp\left(\frac{2\mu}{k_{B}T}\right)\sum_{\mathbf{k},\mathbf{\varkappa},\theta}\frac{2k_{x}k_{y}}{k^{2}}\exp\left(-\frac{2k^{2}+\varkappa^{2}-2k\varkappa \cos\theta}{k_{T}^{2}}\right)\nonumber\\&&=
\frac{\pi^{2} m}{\hbar^{2}}\frac{1}{(2\pi)^{6}}\exp\left(\frac{2\mu}{k_{B}T}\right)\int_{0}^{\infty}kdk\exp\left(-\frac{k^{2}}{k_{T}^{2}}\right)\int_{0}^{2\pi}\sin(2\phi_{\mathbf{k}})d\phi_{\mathbf{k}}\int_{0}^{\infty}\varkappa d\varkappa \exp\left(-\frac{\varkappa^{2}}{k_{T}^{2}}\right)I_{0}\left(\frac{2k\varkappa}{k_{T}^{2}}\right)=0\nonumber\\
\end{eqnarray}
as $\int_{0}^{2\pi}\sin(2\phi_{\mathbf{k}})d\phi_{\mathbf{k}}=0$. Similarly, as $\int_{0}^{2\pi}\sin(2\phi_{\mathbf{k}'})d\phi_{\mathbf{k}'}=0$ one can obtain that $X'_{2a}=0$.
\begin{eqnarray}
&&X'_{2a}=\sum_{\mathbf{k}\mathbf{k}'\mathbf{p}\mathbf{p}'}\delta_{\mathbf{k}+\mathbf{p},\mathbf{k}'+\mathbf{p}'}\delta(\varepsilon_{k}+\varepsilon_{p}-\varepsilon_{k'}-\varepsilon_{p'})\cdot\frac{2k'_{x}k'_{y}}{k'^{2}}f_{\mathbf k}^{0}f_{\mathbf{p}}^{0}=
\frac{m}{\hbar^{2}}\sum_{\mathbf{k},\mathbf{\varkappa},\mathbf{\varkappa}'}\delta(\varkappa^{2}-\varkappa'^{2})\cdot\frac{2k'_{x}k'_{y}}{k'^{2}}f_{\mathbf{k}'+\frac{1}{2}(\mathbf{\varkappa}-\mathbf{\varkappa}')}^{0}f_{\mathbf{k}'-\frac{1}{2}(\mathbf{\varkappa}+\mathbf{\varkappa}')}^{0}\nonumber\\&&=/\theta=\angle(\mathbf{k}',\mathbf{\varkappa}')/
=\frac{\pi m}{\hbar^{2}}\frac{1}{(2\pi)^{2}}\exp\left(\frac{2\mu}{k_{B}T}\right)\sum_{\mathbf{k}',\mathbf{\varkappa},\theta}\frac{2k'_{x}k'_{y}}{k'^{2}}\exp\left(-\frac{2k'^{2}+\varkappa^{2}-2k'\varkappa \cos\theta}{k_{T}^{2}}\right)=\nonumber\\&&=
\frac{\pi^{2} m}{\hbar^{2}}\frac{1}{(2\pi)^{6}}\exp\left(\frac{2\mu}{k_{B}T}\right)\int_{0}^{\infty}k'dk'\exp\left(-\frac{k'^{2}}{k_{T}^{2}}\right)\int_{0}^{2\pi}\sin(2\phi_{\mathbf{k}'})d\phi_{\mathbf{k}'}\int_{0}^{\infty}\varkappa d\varkappa\exp\left(-\frac{\varkappa^{2}}{k_{T}^{2}}\right)I_{0}\left(\frac{2k'\varkappa}{k_{T}^{2}}\right)=0\nonumber\\
\end{eqnarray}

\end{widetext}

\renewcommand{\i}{\ifr}


\begin{thebibliography}{99}
\bibitem{Kolobov2016book} A.~V. Kolobov and J.~Tominaga,
\newblock {Two-Dimensional Transition-Metal Dichalcogenides}.
\newblock Springer International Publishing, 2016.
\bibitem{deotare20232d}
P.~Deotare and Z.~Mi, editors,
\newblock {2D Excitonic Materials and Devices}.
\newblock Elsevier Science, 2023.
\bibitem{Ivchenko2005} E.L. Ivchenko, Optical Spectroscopy of Semiconductor Nanostructures, Alpha Science (2005).
\bibitem{RevModPhys.90.021001} G.~Wang, A.~Chernikov, M.~M. Glazov, T.~F. Heinz, X.~Marie, T.~Amand, and B.~Urbaszek, Colloquium: Excitons in atomically thin transition metal dichalcogenides, Rev. Mod. Phys. {\bf 90}, 021001 (2018).
\bibitem{Geim:2013aa} A.~K. Geim and I.~V. Grigorieva, Van der Waals heterostructures,
Nature {\bf 499}, 419 (2013).
\bibitem{Hillmer1988} H. Hillmer, S. Hansmann, A. Forchel, M. Morohashi, E. Lopez, H.P Meier, and K. Ploog, Two-dimensional exciton transport in GaAs/GaAlAs quantum wells, Appl. Phys. Lett. {\bf 53}(20), 1937 (1988).
\bibitem{Steininger1996} F. Steininger, A. Knorr, T. Stroucken, P. Thomas, and S.W. Koch, Dynamic evolution of spatiotemporally localized electron wave packets in semiconductor quantum wells, Phys. Rev. Lett. {\bf 77}(3), 550 (1996).
\bibitem{Rapaport2004} R. Rapaport, G. Chen, D. Snoke, S.H. Simon, L. Pfeiffer, K. West, Y.Liu, and S. Denev, Charge separation of dense two-dimensional electron-hole gases: mechanism for exciton ring pattern formation, Phys. Rev. Lett. {\bf 92}(11), 117405 (2004).
\bibitem{Kumar2014} N. Kumar, Q. Cui, F. Ceballos, D. He, Y. Wang, and H. Zhao, Exciton diffusion in monolayer and bulk $MoSe_{2}$, Nanoscale {\bf 6}, 4915 (2014).
\bibitem{Kato2016} T. Kato, and T. Kaneko, Transport dynamic of neutral excitons and trions in monolayer $WS_{2}$, ACS Nano {\bf 10}, 9687 (2016).
\bibitem{Baldo2009} M. Baldo, and V. Stojanovic, Excitonic interconnectors, Nat. Photon. {\bf 3}, 558 (2009).
\bibitem{Causin2022} R. Perea-Causin, D. Erkensten, J. M. Fitzgerald, J. J. P. Thompson, R. Rosati, S. Brem, and E. Malic , Exciton optics, dynamics and transport in atomically thin semiconductors, APL Mater. {\bf 10}, 100701 (2022).
%
\bibitem{Malic:2023aa} E.~Malic, R.~Perea-Causin, R.~Rosati, D.~Erkensten, and S.~Brem, Exciton transport in atomically thin semiconductors, Nature Communications {\bf 14}, 3430 (2023).
%
\bibitem{Chernikov:2023ab} A.~Chernikov and M.~M. Glazov, {\em Exciton diffusion in 2D van der Waals
  semiconductors} in 2D Excitonic Materials and Devices ed. by  Parag B. Deotare and  Zetian Mi, Elsevier, 2023.
%
\bibitem{Yuan2017} L. Yuan, T. Wang, T. Zhu, M. Zhou, and L. Huang, Exciton dynamics, transport and annihilation in atomically thin two-dimensional semiconductors, J. Phys. Chem. Lett. {\bf 8}, 3371 (2017).
\bibitem{Vlaming2013} S.M. Vlaming, V.A. Malyshev, A. Eisfeld, and J. Knoester, Subdiffusive exciton motion in systems with heavy-tailed disorder, J. Chem Phys. {\bf 138}, 214316 (2013).
\bibitem{Kurilovich2020} A.A. Kurilovich, V.N. Mantsevich, K.J. Stevenson, A.V. Chechkin, and V.V. Palyulin, Complex diffusion-based kinetics of photoluminescence in semiconductor nanoplatelets, Phys. Chem. Chem. Phys. {\bf 22}, 24686 (2020).
\bibitem{Kurilovich2022} A.A. Kurilovich, V.N. Mantsevich, Y. Mardoukhi, K.J. Stevenson, A.V. Chechkin, and V.V. Palyulin, Non-Markovian diffusion of excitons in layered perovskites and transition metal dichalcogenides, Phys. Chem. Chem. Phys. {\bf 24}, 13941 (2022).
\bibitem{Lee2015} E. M.Y. Lee, and W.A. Tisdale, Determination of exciton diffusion length by transient photoluminescence quenching and its application to quantum dot films, J. Phys. Chem. C {\bf 119}, 9005 (2015).
\bibitem{Uddin2020} S.Z. Uddin, H. Kim, M. Lorenzon, M. Yeh, D.-H. Lien, E.S. Bernard, H. Htoon, A. Weber-Bargioni, and A. Javey, Neutral exciton diffusion in monolayer $MoS_{2}$, ACS Nano {\bf 14}, 13433 (2020).
\bibitem{Cheng2021} G. Cheng, B. Li, Z. Jin, M. Zhang, and J. Wang, Observation of diffusion and drift of negative trions in monolayer $WS_{2}$, Nano Lett. {\bf 21}, 6314 (2021).
\bibitem{Wagner2023} K. Wagner, Z.A. Iakovlev, J.D. Ziegler, M. Cuccu, T. Taniguchi, K. Watanabe, M.M. Glazov, and A. Chernikov, Diffusion of excitons in a two-dimensional Fermi sea of free charges, Nano Lett. {\bf 23}, 4708 (2023).
\bibitem{Glazov2019} M.M. Glazov, Phonon wind and drag of excitons in monolayer semiconductors, Phys. Rev. B {\bf 100}, 045426 (2019).
\bibitem{Shklovskii1984} B.I. Shklovskii, and A.L. Efros, Electronic properties of doped semiconductors, Vol. 45 of Springer Series in Solid-State Sciences, Springer Berlin Heidelberg (1984).
\bibitem{Gantmakher1987} V.F. Gantmakher, and Y.B. Levinson, Carrier scattering in metals and semiconductors, North Holland Physics Publishing, Amsterdam (1987).
\bibitem{Zipfel2020} J. Zipfel, M. Kulig, R. Perea-Causin, S. Brem, J.D. Ziegler, R. Rosati, T. Taniguchi, K. Watanabe, M.M. Glazov, E. Malic, and A. Chernikov, Exciton diffusion in monolayer semiconductors with suppressed disorder, Phys. Rev. B {\bf 101}, 115430 (2020).
\bibitem{Choi2023} J. Choi, J. Embly, D.D. Blach, R. Perea-Causin, D. Erkensten, D.S. Kim, L. Yuan, W.Y. Yoon, T. Taniguchi, K. Watanabe, K. Ueno, E. Tutic, S. Brem, E. Malic, X. Li, and L. Huang, Fermi pressure and Coulomb repulsion driven rapid hot plasma expansion in a van der Waals heterostructure, Nano Lett. {\bf 23}, 4399 (2023).
\bibitem{Glazov2020} M.M. Glazov, Quantum interference effect on exciton transport in monolayer semiconductors, Phys. Rev. Lett. {\bf 124}, 166802 (2020).
\bibitem{Kenkre1983} V.M. Kenkre, Master equation techniques for exciton motion, relaxation, capture, and annihilation, J. Stat. Phys. {\bf 30}, 293 (1983).
\bibitem{Heijs2005} D.J. Heijs, V.A. Malyshev, and J. Knoester, Decoherence of excitons in multichromophore systems: thermal line broadening and destruction of superradiant emission, Phys. Rev. Lett. {\bf 95}, 177402 (2005).
\bibitem{Akselrod2014_1} G.M. Akselrod, F. Prins, L.V. Poulikakos, E. M.Y. Lee, M.C. Weidman, A.J. Mork, A.P. Willard, V. Bulovich, and W.A. Tisdale, Subdiffusive exciton transport in quantum dot solids, Nano Lett. {\bf 14}, 3556 (2014).
\bibitem{Miyazaki2012} J. Miyazaki, and S. Kinoshita, Site-selective spectroscopic study on the dynamics of exciton hopping in an array of inhomogeneously broadened quantum dots, Phys. Rev. B {\bf 86}, 035303 (2012).
\bibitem{Wietek2024} E. Wietek, M. Florian, J. Goser, T. Taniguchi, K. Watanabe, A. Hogele, M. M. Glazov, A. Steinhoff, and A. Chernikov, Nonlinear and negative effective diffusivity of interlayer excitons
in Moire-free heterobilayers, Phys. Rev. Lett. {\bf 132}, 016202 (2024).
\bibitem{Kurilovich2023} A.A. Kurilovich, V.N. Mantsevich, A.V. Chechkin, and V.V. Palyulin, Negative diffusion of excitons in quasi-twodimensional systems, Phys. Chem. Chem. Phys. {\bf 26}, 922 (2024).



%
\bibitem{Qiu:2021to} D.~Y. Qiu, G.~Cohen, D.~Novichkova, and S.~Refaely-Abramson, Signatures of dimensionality and symmetry in exciton band structure:   Consequences for exciton dynamics and transport,  Nano Letters {\bf 21}, 7644 (2021).
%
\bibitem{Fogler2014} M.~M. Fogler, L.~V. Butov, and K.~S. Novoselov, High-temperature superfluidity with indirect excitons in van der  Waals heterostructures, Nature Communications {\bf 5}, 4555 (2014).
%
\bibitem{GlazovSuris_2021} M.~M. Glazov and R.~A. Suris, Collective states of excitons in semiconductors, Physics-Uspekhi {\bf 63}, 1051 (2021).
%
\bibitem{Aguila2023} A.G. del Aguila, Y.R. Wong, I. Wadgaonkar, A. Fieramosca, X. Liu, K. Vaklinova, S. Dal Forno, T. Thu Ha Do, H.Y. Wei, K. Watanabe, T. Taniguchi, K.S. Novoselov, M. Koperski, M. Battiato, and Q. Xiong, Ultrafast exciton fluid flow in an atomically thin $MoS_2$ semiconductor, Nat. Nanotechnol. {\bf 18}, 1012 (2023).
%
\bibitem{Glazov:2023aa} M.~M. Glazov, Excitons in atomically thin materials flow faster than they fly, Nature Nanotechnology {\bf 18}, 972 (2023).
%
\bibitem{glazov2024ultrafastexcitontransportvan} M.~M. Glazov and R.~A. Suris, Ultrafast exciton transport in van der waals heterostructures, arXiv:2403.19571 (2024); ZhETF {\bf 166}, 20 (2024).
%

\bibitem{Gurzhi_1968} R.~N. Gurzhi, Hydrodynamic effects in solids at low temperatures, Soviet Physics Uspekhi {\bf 11}, 255 (1968).



\bibitem{PhysRevLett.106.256804} A.~V. Andreev, S.~A. Kivelson, and B.~Spivak, Hydrodynamic description of transport in strongly correlated electron systems, Phys. Rev. Lett. {\bf 106}, 256804 (2011).



\bibitem{PhysRevLett.117.166601} P.~S. Alekseev, Negative magnetoresistance in viscous flow of two-dimensional electrons, Phys. Rev. Lett. {\bf 117}, 166601 (2016).



\bibitem{PhysRevB.51.13389} M.~J.~M. de~Jong and L.~W. Molenkamp, Hydrodynamic electron flow in high-mobility wires, Phys. Rev. B {\bf 51}, 13389 (1995).




\bibitem{Bandurin1055} D.~A. Bandurin, I.~Torre, R.~K. Kumar, M.~Ben~Shalom, A.~Tomadin, A.~Principi, G.~H. Auton, E.~Khestanova, K.~S. Novoselov, I.~V. Grigorieva, L.~A. Ponomarenko, A.~K. Geim, and M.~Polini, Negative local resistance caused by viscous electron backflow in  graphene, Science {\bf 351}, 1055 (2016).



\bibitem{Crossno:2016aa} J.~Crossno, J.~K. Shi, K.~Wang, X.~Liu, A.~Harzheim, A.~Lucas, S.~Sachdev,
  P.~Kim, T.~Taniguchi, K.~Watanabe, T.~A. Ohki, and K.~C. Fong, Observation of the Dirac fluid and the breakdown of the  Wiedemann-Franz law in graphene, Science {\bf 351}, 1058 (2016).



\bibitem{Moll1061} P.~J.~W. Moll, P.~Kushwaha, N.~Nandi, B.~Schmidt, and A.~P. Mackenzie, Evidence for hydrodynamic electron flow in PdCoO$_2$, Science {\bf 351}, 1061 (2016).




\bibitem{Levitov:2016aa} L.~Levitov and G.~Falkovich, Electron viscosity, current vortices and negative nonlocal resistance in graphene, Nature Physics {\bf 12}, 672 (2016).



\bibitem{Gusev:2018tg} G.~M. Gusev, A.~D. Levin, E.~V. Levinson, and A.~K. Bakarov, Viscous electron flow in mesoscopic two-dimensional electron gas, AIP Advances {\bf 8}, 025318 (2018).



\bibitem{PhysRevLett.128.136801} Y.~A. Pusep, M.~D. Teodoro, V.~Laurindo, E.~R. Cardozo~de Oliveira, G.~M. Gusev, and A.~K. Bakarov, Diffusion of photoexcited holes in a viscous electron fluid, Phys. Rev. Lett. {\bf 128}, 136801 (2022).

\bibitem{Narozhny:2022ud} B.~N. Narozhny, Hydrodynamic approach to two-dimensional electron systems, La Rivista del Nuovo Cimento {\bf 45} 661, (2022).

\bibitem{ll10_eng} L.~Landau and E.~Lifshitz, {\em Physical Kinetics}, Butterworth-Heinemann, Oxford, 1981.

\bibitem{Irving:1950aa} J.~H. Irving and J.~G. Kirkwood,
The statistical mechanical theory of transport processes. IV. The equations of hydrodynamics, J. Chem. Phys. {\bf 18}, 817 (1950).



\bibitem{ll6_eng} L.~D. Landau and E.~M. Lifshitz, {\em Fluid Mechanics}, Pergamon Press, 1987.

\bibitem{PhysRevLett.127.076801} K.~Wagner, J.~Zipfel, R.~Rosati, E.~Wietek, J.~D. Ziegler, S.~Brem, R.~Perea-Causin, T.~Taniguchi, K.~Watanabe, M.~M. Glazov, E.~Malic, and A.~Chernikov, Nonclassical exciton diffusion in monolayer {WSe}$_{2}$, Phys. Rev. Lett. {\bf 127}, 076801 (2021).





\bibitem{Kaasbjerg2012} K. Kaasbjerg, K.S. Thygesen, and K.W. Jacobsen, Phonon limited mobility in n-type single-layer $MoS_2$ from first
principles, Phys. Rev. B {\bf 85}, 115317 (2012).
\bibitem{Shree2018} S. Shree, M. Semina, C. Robert, B. Han, T. Amand, A. Balocchi, M. Manca, E. Courtade, X. Marie, T. Taniguchi, K. Watanabe, M.M. Glazov, and B. Urbaszek, Observation of exciton-phonon coupling in $MoSe_2$ monolayers, Phys. Rev. B {\bf 98}, 035302 (2018).

\bibitem{PhysRevLett.105.070401} A.~Filinov, N.~V. Prokof'ev, and M.~Bonitz, {Berezinskii-Kosterlitz-Thouless} transition in two-dimensional
  dipole systems,  Phys. Rev. Lett. {\bf 105}, 070401 (2010).



\bibitem{Shahnazaryan2017} V. Shahnazaryan, I. Iorsh, I.A. Shelykh, and O. Kyriienko, Exciton-exciton interaction in transition-metal dichalcogenide monolayers, Phys. Rev. B {\bf 96}, 115409 (2017).

\bibitem{Goryca2019} M. Goryca, J. Li, A. V. Stier, T. Taniguchi, K. Watanabe, E. Courtade, S. Shree, C. Robert, B. Urbaszek, X. Marie, and S. A. Crooker, Revealing exciton masses and dielectric properties
of monolayer semiconductors with high magnetic fields, Nat. Comm. {\bf 10}, 4172 (2019).

\bibitem{Han2018} B. Han, C. Robert, E. Courtade, M. Manca, S. Shree, T. Amand, P. Renucci, T. Taniguchi, K. Watanabe, X. Marie, L. E. Golub, M. M. Glazov, and B. Urbaszek, Exciton States in Monolayer $MoSe_2$ and $MoTe_2$ Probed by Upconversion Spectroscopy, Phys. Rev. X,  {\bf 8}, 031073 (2018).

\bibitem{Kormanyos2015} A. Kormányos, G. Burkard, M. Gmitra, J. Fabian,  V. Zólyomi,  N. D. Drummond, and V. Fal’ko, k-p theory for two-dimensional transition metal dichalcogenide semiconductors, 2D Materials,  {\bf 2}, 022001 (2015).

\bibitem{Berkelbach2013} T.C. Berkelbach, M.S. Hybertsen, and D.R. Reichman, Theory of neutral and charged excitons in monolayer transition metal dichalcogenides, Phys. Rev. B {\bf 88}, 045318 (2013).
\bibitem{Jin2014} Z. Jin, X. Li, J.T. Mullen, and K.W. Kim, Intrinsic transport properties of electrons and holes in monolayer transition-metal dichalcogenides, Phys. Rev. B {\bf 90}, 045422 (2014).
\bibitem{Erkensten2021} D. Erkensten, S. Brem, and E. Malic, Exciton-exciton interaction in transition metal dichalcogenide monolayers and van der Waals heterostructures, Phys. Rev. B {\bf 103}(4), 2469 (2021).
\bibitem{Steinhoff2023} A. Steinhoff, Edith Wietek, M. Florian, T. Schulz, T. Taniguchi, K. Watanabe, S. Zhao, A. H\"{o}gele, F. Jahnke, and A. Chernikov, Exciton-exciton interactions in van der Waals heterobilayers, Phys. Rev. X {\bf 14}, 031025 (2024).
\bibitem{Tagarelli2023} F. Tagarelli, E. Lopriore, D. Erkensten,  R. Perea-Causin,  S. Brem,  J. Hagel, Z. Sun, G. Pasquale, K. Watanabe, T. Taniguchi, E. Malic, and A. Kis, Electrical control of hybrid exciton transport in a van der Waals heterostructure, Nat. Photonics {\bf 17}(7), 615 (2023).
\bibitem{Sun2022} Z. Sun, A. Ciarrocchi, F. Tagarelli, J. F. Gonzalez Marin, K. Watanabe, T. Taniguchi, and A. Kis, Excitonic transport driven by repulsive dipolar interaction in a van der Waals heterostructure, Nature Photonics {\bf 16}, 79 (2022).
\bibitem{Kulig2018} Marvin Kulig, Jonas Zipfel, Philipp Nagler, Sofia Blanter, Christian Sch\"uller, Tobias Korn, Nicola Paradiso, Mikhail M. Glazov, and Alexey Chernikov, Exciton Diffusion and Halo Effects in Monolayer Semiconductors, Phys. Rev. Lett. {\bf 120}, 207401 (2018)
\bibitem{Kuznetsov:2020aa}
E.~A. Kuznetsov and M.~Y. Kagan, Semiclassical expansion of quantum gases into a vacuum, Theoretical and Mathematical Physics {\bf 202}, 399 (2020).
\bibitem{Yuan2020} L. Yuan, B. Zheng, J. Kunstmann, T. Brumme, A. B. Kuc, C. Ma, S. Deng, D. Blach, A. Pan, and L. Huang, Twist angle-dependent interlayer exciton diffusion in $WS_2-WSe_2$ heterobilayers, Nature Materials {\bf 19}, 617 (2020).
\bibitem{Erkensten2022} D. Erkensten, S. Brem, R. Perea-Causin, and E. Malic, Microscopic origin of anomalous interlayer exciton transport in van der Waals heterostructures, Phys. Rev. Matter. {\bf 6}(9), 094006 (2022).
\bibitem{PhysRevB.106.235305} M.~M. Glazov and L.~E. Golub, Spin and valley {Hall} effects induced by asymmetric interparticle scattering, Phys. Rev. B {\bf 106}, 235305 (2022).







\end{thebibliography}
\end{document}